\documentclass[aps,prl,twocolumn,showpacs,byrevtex,amsmath,amssymb]{revtex4}

\usepackage{graphicx}

\begin{document}

\title{Oscillatory driven colloidal binary mixtures: axial segregation versus laning}

\author{Adam Wysocki and Hartmut L\"owen}

\affiliation{Institut f\"ur Theoretische Physik II: Weiche Materie, 
Heinrich-Heine-Universit\"at D\"usseldorf, Universit\"atsstrasse 1, 
D-40225 D\"usseldorf, Germany}

\date{\today}

\begin{abstract}

Using Brownian dynamics computer simulations we show that binary mixtures of colloids
driven in opposite directions by an oscillating external field 
exhibit axial segregation in sheets perpendicular to the drive 
direction. The segregation effect is stable only in a finite window of 
oscillation frequencies and driving strengths and is taken over by lane formation 
in direction of the driving field if the driving force is increased. In the crossover regime,
bands tilted relative to the drive direction are observed.
Possible experiments to verify the axial segregation are discussed.

\end{abstract}


\maketitle

Phase transitions in driven systems are qualitatively different from their 
equilibrium counterparts and reveal a wealth of novel instabilities and 
pattern formations induced by nonequilibrium conditions \cite{schmittmann}. 
In particular, it is very intriguing to resolve and follow nonequilibrium 
dynamics on the particle level which is possible for granular grains 
\cite{aranson2006,ottino2000}, mesoscopic colloidal suspensions 
\cite{loewen2001} or dusty plasmas driven by an external field \cite{suetterlin2009}.

In the colloidal context, recent investigations have focussed on binary 
mixtures which are driven by a constant but species-dependent force. 
The latter can be realized by gravity \cite{royall2007,okubo2008} 
or by an electric field \cite{leunissen2005}. 
If the drive is strong enough, lanes of particles driven alike are formed 
in direction of the driving field  \cite{dzubiella2002a}. 

In this paper we simulate a model for a colloidal mixture with an 
time-dependent oscillatory drive \footnote[1]{Recent investigations with oscillatory 
drives have revealed quite rich non-equilibrium transitions including e.g. drive-induced 
ergodicity breaking, see e.g. \cite{corte2008,mangan2008}.}. We show that this induces an axial segregation, i.e.\ an
ordering in sheets normal to the driving field (rather than parallel as for 
laning). While such an axial segregation of two particle species is observed in 
shaken or vibrated granular systems \cite{mullin2000,sanchez2004,Ciamarra2005,Ciamarra2007}, 
it has not yet been described for colloidal suspensions whose dynamics 
is dominated by overdamped Brownian motion. 
This makes the colloidal dynamics different from that of granular systems 
where inelastic collisions play an important role. 
At fixed density, we map out the whole non-equilibrium steady-state diagram
over a large range of frequencies and driving amplitudes and find
a rich phase diagram topology. By increasing the magnitude of the driving force or 
decreasing the oscillation frequency, the system gets back either to a disordered state
or to lane formation with an intermediate regime where mixing of segregation and laning
as well as tilted lanes occur. Decreasing the magnitude of the driving
or increasing the oscillation frequency, on the other hand, always yields a disordered 
steady state. We then study the relaxational dynamics into the steady states and show
that first laning occurs and if at all axial segregation occurs on a larger time scale.
Finally we discuss the anisotropy of the mean-square-displacements
in the different steady states as dynamical consequences of their different structure.
\begin{figure}
\includegraphics[width=8cm]{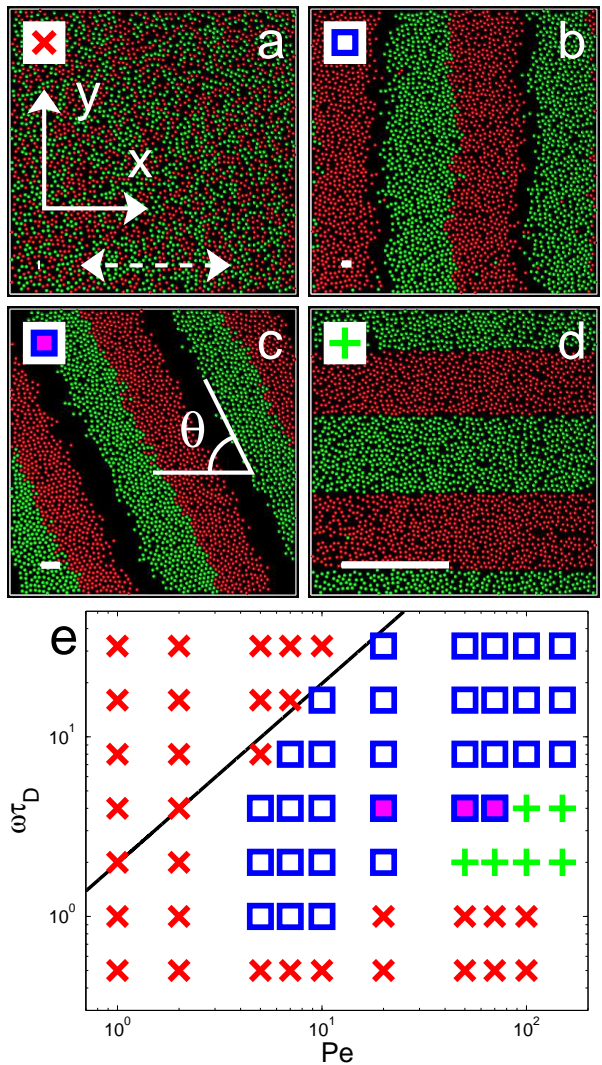}
\caption{{\bf a-d} BD simulation snapshots for fixed $\omega\tau_D=4$ but different 
Peclet numbers $Pe$ after $10^4$ periods starting from a fully mixed configuration. 
Symbols in the left corner correspond to symbols used in ({\bf e}). 
In {\bf a} the coordinate frame is shown and the direction of the driving field is indicated 
by the broken arrow.
In {\bf a-d} the length of the solid bars correspond to the amplitude of a free particle
driven without noise in the external field.
For small $Pe$, we observe a disordered state ({\bf a}), for intermediate $Pe$ 
colloids segregate into stripes oriented perpendicular or tilted to the direction 
of the oscillating force ({\bf b-c}), on the other hand, for high $Pe$, lanes are formed 
parallel to the direction of the oscillatory force ({\bf d}). 
Parameters are $\omega\tau_{D}=4$ and $Pe=2, 10, 20, 110$ from {\bf a} to {\bf d}.
{\bf e} Nonequillibrium steady state phase diagram for fixed area fraction $\phi=0.4$.
The solid line describes the simple theoretical estimate of the 
disordered-to-segregated phase boundary.}
\label{figPhasediagram} 
\end{figure} 

In order to simulate the dynamics of a colloidal suspension 
we utilise a two dimensional (2D) Brownian dynamics (BD) simulation with neglected
 hydrodynamic interactions between the particles. 
In particular, we consider an equimolar binary mixture of $N=2864$ hard disks of diameter $\sigma$.
In Brownian dynamics, the time evolution of the individual colloids is governed by 
 overdamped Langevin equation with thermal noise \cite{allen1991}. 
The particles of sort $A$ and $B$ are subjected to an oscillatory inversely 
phased force ${\bf f}^{A}(t)=-{\bf f}^{B}(t)=f_0\sin{(\omega t)}{\bf e}_{x}$, 
where $\omega$ is the driving frequency and $f_0$ the driving strength.
A single  particle will follow a  noise-averaged trajectory which is sinusoidal in time
with an amplitude $f_0/\xi\omega$.

We simulate a quadratic system of size $L/\sigma=75$ with periodic boundary conditions 
at a total area fraction $\phi=N\pi\sigma^2/(4L^2)=0.4$. As a measure for the driving strength,
we define the  Peclet number 
$Pe=\tau_{D}/\tau_{d}$ as the ratio between the time $\tau_D=\sigma^2/(4D)$ 
 it takes a colloid to diffuse its own radius and the time $\tau_{d}=\sigma/(2v_{d})$ 
it takes to drift the same distance under the action of a constant external force
$f_0$. Here $D=k_BT/\xi$ denotes the short-time diffusion constant
where $k_B T$ is the thermal energy, $\xi$ 
the friction constant and $v_d = f_0/\xi$ is the maximal drift velocity.

The initial configuration of the BD simulation 
is  an equilibrated fully mixed configuration as realized in the absence of the drive. 
At time $t=0$, the oscillatory force is turned on 
instantaneously and the evolution of the system is followed by using a finite time-step
algorithm \cite{wysocki2004} up to about $10^5$ oscillation periods. Depending on the Peclet number 
 $Pe$ and the reduced frequency $\omega\tau_{D}$, the system ends up in different steady states. 

The nonequilibrium steady-state phase diagram is shown in  Fig.\ref{figPhasediagram}
spanning two decades in both driving frequency $\omega\tau_{D}$ and strength $Pe$.
Cooresponding movies showing the development from the initial to the steady state are avaibale
in \footnote[2]{Movies of the simulation can be viewed at http://www2.thphy.uni-duesseldorf.de/\symbol{126}adam/}. 
Let us discuss the topology of the phase diagram step by step: 
For very low driving strengths, the external drive is not strong enough to generate
order in the completely mixed equilibrium state leading to a {\it disordered\/} steady-state which is
depicted in a typical snapshot in Fig.\ref{figPhasediagram} \textbf{a}.
A disordered state is rendered as crosses in the full phase diagram shown 
in Fig.\ref{figPhasediagram}\textbf{e} such that the region of low Peclet numbers
involves a disordered steady state. As visible
in  Fig.\ref{figPhasediagram} \textbf{a}, the disordered state can possess an 
intrinsic finite correlation length
as set by colliding clusters which is larger than the interparticle distance
but these structures
do not span the whole simulation box.
For very high frequencies, on the other hand,
a free particle would just perform oscillations with a very small amplitude
due to the drive. This amplitude  $v_{d}/\omega$ is indicated
as an inset  in the corresponding simulation snapshots
in Fig.\ref{figPhasediagram} \textbf{a}-\textbf{d}.
At high frequencies, an ensemble of many  particles, 
is just rattling over a small distance and is thus not perturbing much the
disordered equilibrium state. Conversely, at very low frequencies, {\it lane formation\/}
similar to that observed earlier for a static or slowly oscillating drive \cite{dzubiella2002a} occurs
where the direction of the lanes coincides with the drive direction.
A corresponding simulation snapshot is given in Fig.\ref{figPhasediagram} \textbf{d}.
In a region of finite frequencies and Peclet numbers,
 bounded by  disordered and laned steady states, axial segregation shows up, 
where the stripes are oriented perpendicular to the drive direction,
see the corresponding snapshot in Fig.\ref{figPhasediagram} \textbf{b}.
Here the particles driven alike perform collectively oscillations as induced by the
external drive but collide periodically with an opposing band of opposite particles.
In Fig.\ref{figPhasediagram} \textbf{b}, a colliding interface is seen in the middle
of the snapshot while voids in the two other interface indicate that the bands had just
been driven away from this position. We call this stripe formation 
perpendicular to the driving field {\it axial segregation}. Finally, in an intermediate region
between axial segregation and laning, some steady states are observed with {\it tilted\/} bands,
see e.g. the snapshot in Fig.\ref{figPhasediagram} \textbf{c}, where a general 
tilt angle $\theta$ relative to the drive direction is observed. However, as is discussed in
detail below, the major part of the steady states are still either segregated or laned.

There are two {\it reentrant effects\/} of the disordered phase 
in the phase diagram shown in Fig.\ref{figPhasediagram} \textbf{e}:
first by increasing the Peclet number at fixed frequency $\omega\tau_{D}\approx 1$,
the steady state transforms from the disordered one into the axially segregated one and
back to the disordered one. Second, now for increasing frequency at fixed
Peclet number $Pe<5 \approx 10$, the disordered phase is taken over by 
the axially segregated phase and then appears again.

A simple quantitative criterion for the disordered-to-segregated transition at high frequencies
can be derived by the intuitive argument that the oscillating amplitude $v_{d}/\omega$ must
exceed the mean interparticle distance $\sigma(\sqrt{\phi_{max}/\phi}-1)$ in order to
induce a noticable segregation effect. The analytical estimate of this phase boundary line
is shown as a solid line in  Fig.\ref{figPhasediagram} \textbf{e} and agrees well with the
simulation data at intermediate and high Peclet numbers.
\begin{figure}
\includegraphics[width=8cm]{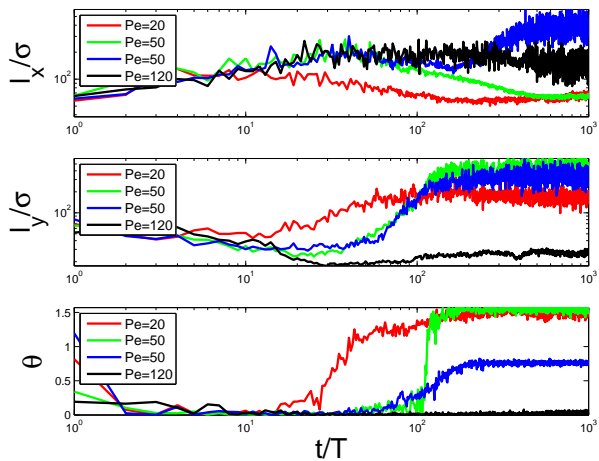}
\caption{Time evolution of a typical length scale parallel ({\bf a}) and perpendicular ({\bf b})  
to the direction of the oscillatory force $l_{x}/\sigma$ and $l_{y}/\sigma$, 
as well as of the orientation $\theta$ of a typical structure ({\bf c}). In the BD simulations 
the angular frequency is fixed $\omega\tau_{D}=4$ and the Peclet numbers are $Pe=20,50,120$.
For $Pe=50$, both a tilted and segregated phase are shown.}
\label{figLengthAngle} 
\end{figure}

We now turn to the dynamics of how the different steady states are reached 
after turning on the driving field.
For this purpose we stroboscopically determine the location 
of the colloids after each drive period $T$ (at a time where all external forces vanish) 
and calculate the partial static structure factor 
$S({\bf k},nT)=2/N|\sum_{i=1}^{N/2}e^{-i{\bf k}{\bf r}_i}|^2$ in a range 
of wavenumbers $k_{\alpha}\in[-k_{max},k_{max}]$ where $\alpha\in\{x,y\}$ and $k_{max}=2\pi/\sigma$. 
Here ${\bf r}_i$ with $i\in\{1,\ldots,N/2\}$ denote the positions of all particles of the same species. 
From the structure factor we get the typical length scale of the system via a first moment
defined as
\begin{equation}
\frac{2\pi}{l_{\alpha}(nT)}=\frac{\int_{-k_{max}}^{k_{max}}|k_{\alpha}|S({\bf e}_{\alpha}{\bf k},nT)dk_{\alpha}}{\int_{-k_{max}}^{k_{max}} S({\bf e}_{\alpha}{\bf k},nT)dk_{\alpha}}
\end{equation}
To obtain the orientation of a typical structure in the system with respect to the direction 
of the oscillatory force we calculate the tensor
\begin{equation}
{\bf S}=\iint_{-k_{max}}^{k_{max}}S({\bf k},nT)\left[({\bf k}{\bf k}){\bf I}-{\bf k}\otimes{\bf k}\right]d{\bf k}
\end{equation}
(where ${\bf I}$ denotes the unit tensor and $\otimes$ the dyadic product)
and define $\theta$ as the angle between ${\bf e}_x$ and the eigenvector $\hat{{\bf e}}_1$ corresponding 
to the biggest eigenvalue of ${\bf S}$. The angle $\theta$ serves as an "order parameter"
of the observed structures: in fact, $\theta=0$ points to a state of lanes while 
$\theta=\pi/2$ corresponds to axial segregated structures and
 $0<\theta<\pi/2$ indicates a tilted state.

The time evolution of the two length scales $l_x$ and $l_y$ as well as the angle $\theta$
is shown in Fig.\ref{figLengthAngle} at fixed driving frequency $\omega\tau_{D}=4$ for three
different Peclet numbers and four different end states two 
two of which are segregated end-state and one is laned and tilted. 
 Strikingly, the system first develops structures
along the drive which is persistent over more than a decade in time. It is only after this induction time that the system 
goes into the final segregated state. This clearly shows that
the formation of the axially segregated steady state is a highly collective phenomenon which cannot
by accounted for by a local instability analysis. Laning, on the other hand, can be
viewed as a local instability \cite{chakrabarti2003,chakrabarti2004}.

Let us now discuss the tilted state in some more detail: in the range $30<Pe<80$ some 
steady states possess a non-trivial winding around our "toroidal" system 
(which is dictated by the periodic boundary conditions in the two dimensional system), 
i.e. $\theta=\arctan(n/m)$ with natural numbers $n,m$ and more precisely we observe $n/m=1/2, 1/1, 2/1$. 
We remark that similar  steady states were also found in 
constant driven diffusive lattics gases \cite{bassler1993,schmittmann} and in binary mixtures which are driven by nonparallel 
external forces \cite{dzubiella2002b}. We observe that with increasing $Pe$ the 
probability to end in a tilted  steady state 
with $\theta=\pi/4$ increases and gets nearly equal to that of an axially segregated ($\theta=\pi/2$) state
for $Pe=60$. In the range $50<Pe<60$ there exists even a finite probability for a third steady state 
$\theta=\arctan(2)$. On average the tilted state is reached after $10^2-10^3$ periods and was always 
stable during further $10^5$ periods. We checked that doubling the system
size yields again a tilted state  with same angle $\theta$. Though this clearly shows that
tilting is persistent, we cannot exclude at this stage that the tilting is an artifical
finite-size effect stemming from the toroidal
boundary conditions. However, one can
 speculate \cite{bassler1993} that the lifetime 
of the tilted states diverges with system size.  If so, this
would be an example for multistability in a finite 
region of the $Pe-\omega\tau_D$ space as a consequence of nonequillibrium dynamics \cite{bennett1985}.
\begin{figure}
\includegraphics[width=6.45cm]{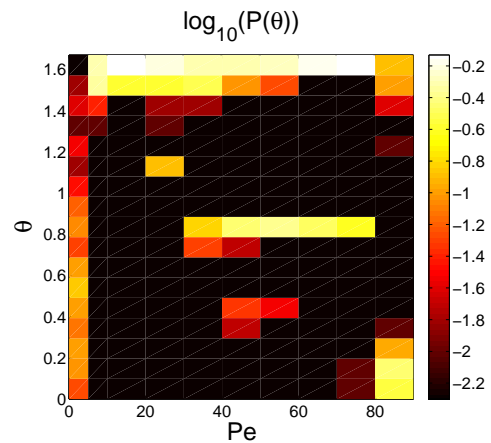}
\caption{Distribution of the orientation $\theta$ of the steady state structures starting from 
a fully mixed configuration after a simulation time of $10^4T$. For every Peclet number 200 independent 
simulations have been performed. The angular frequency is fixed to $\omega\tau_{D}=4$.}
\label{figDistribution} 
\end{figure}

\begin{figure}
\includegraphics[width=8cm]{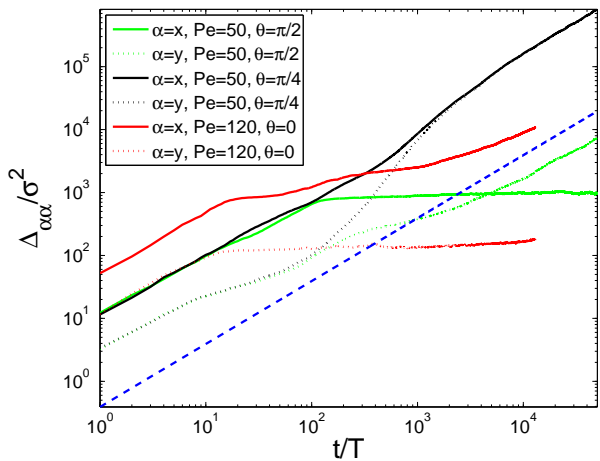}
\caption{Log-log plot of the mean square displacement $\Delta_{xx}$ (solid line), 
$\Delta_{yy}$ (dotted line) for fixed $\omega\tau_{D}=4$. 
Green and black lines are for $Pe=50$ and orientations $\theta=\pi/2$ and $\theta=\pi/4$ respectively. 
Red lines represent $\Delta_{\alpha\alpha}$ for $Pe=120$ and $\theta=0$. The long-dashed blue line correspond 
to the mean square displacement of a free colloid.} 
\label{figDisplFieldMsd} 
\end{figure}

We finally monitor the mean-square-displacement as a function of time $t\equiv nT$ 
in the region of multistability and define the corresponding displacement correlations:
\begin{equation}
\Delta_{\alpha\beta}(t)=1/N\sum_{i=1}^{N}(\Delta{\bf r}_i(t)\cdot{\bf e}_{\alpha})(\Delta{\bf r}_i(t)\cdot {\bf e}_{\beta})
\end{equation}
where $\Delta{\bf r}_i(t)={\bf r}_{i}(t)-{\bf r}_{i}(0)$ and $\alpha,\beta\in\{x,y\}$. 
More precisely we consider 
systems which end up in one of the three steady states with $\theta=0, \pi/4, \pi/2$. 
The two cases $\pi/4, \pi/2$ are calculated at the same Peclet number $Pe=50$.
Results for $\Delta_{\alpha\beta}(t)$ are presented in Fig.\ref{figDisplFieldMsd} 
on a double-logarithmic scale. We conclude that for small times the motion is diffusive
while for long times it is either diffusive or bounded. A quasi-bounded motion arises if the
mean-square-displacement is taken for motion perpendicular to the structures formed.
In comparing the axially segregated state with the tilted state,
the mean square displacement of the two systems is indistinguishable
during the first $10^2$ periods, 
but enhanced compared to a free colloid due to enormous number of interparticle collisions which
increase the particle mobility. Moreover, $\Delta_{xx} > \Delta_{yy}$ for small times due to  initial local
laning.
For $n>10^2$ and $\theta=\pi/2$, particles can unhindered diffuse in 
$y$ direction and are trapped in $x$ direction ($\Delta_{xx}$ nearly constant), while for $\theta=\pi/4$ 
the particles diffuse equally in $x$ and $y$ direction. Comparable to a finite mass current along the 
tilted interface in driven lattics gases \cite{bassler1993}, we notice that for $\theta=\pi/4$ the 
offdiagonal part of the MSD does not vanish $\Delta_{xy}\neq0$ which is  a consequence of a
shear-flow-like displacement field. This is opposed to $\theta=0, \pi/2$ where  
$\Delta_{xy}$ vanishes. 

In conclusion, we have shown that two species of oscillatory driven colloids
segregate into stripes perpendicular to the drive for appropriate frequencies and driving strengths.
We have performed additional simulations which show that axial segregation is stable in various situations:
i) for softer interparticle interactions,
ii) in three spatial dimensions 
where the stripes are two-dimensional sheets of finite width,
iii) in slit-geometry confinement when the driving is perpendicular or parallel to the slit,
iv) if hydrodynamic interactions mediated by a solvent flow are included in the simulation
and the physical volume fraction is small.
The disordered state can be reentrant for increasing the driving frequency and strength.
For increasing strength there is a transition towards lane formation
delineated by a mixed situation of laning and segregation where tilted stripes do also occur.

In principle, it is possible to verify the segregation effect in real-space experiments 
on driven binary colloidal mixtures. Experimental realisations are superparamagnetic colloids 
driven by a gradient in a magnetic field \cite{erbe2008}, oppositely charged 
colloidal mixtures exposed to an alternating electric field \cite{leunissen2005}, 
colloids driven by gravity \cite{erbe2008,royall2007} in a rotating cell
 or  colloidal mixtures driven by dielectrophoretic force in a nonuniform AC electric 
field \cite{hoffman2008,zhao2008}. In particular we think that
  opposite charged colloids in an AC electric field
are a very promising realization to see axial segregation \footnote[3]{The jamming in bands perpendicular 
to the field discussed in Ref.\ \cite{leunissen2005}, however, is different from segregation since a
DC electric field was applied.}. Since hydrodynamic interactions are strongly screened
in this case \cite{rex2008}, our model suitably generalized to three spatial 
dimensions and Yukawa pair interactions  should be appropriate
to describe this system. 

It would be interesting to explore higher densities where freezing occurs \cite{dzubiella2002b}
and to investigate the solvent flow field which might strike back to the pattern 
formation at intermediate physical volume fractions.
Finally, it is challenging to describe axial 
segregation by a microscopic theory based on the Smoluchowski equation 
where either dynamical density functional 
theories \cite{chakrabarti2003,chakrabarti2004} or mode-coupling approaches \cite{brader2007} could be used. 
We leave this important study for the future.

\acknowledgements

We thank R. Zia, A. van Blaaderen, A. Imhof, P. Royall, M. P. Ciamarra and T. Vissers for helpful discussions. 
We acknowledge ZIM D\"usseldorf and J\"ulich Supercomputing Center (SoftComp Cluster) for 
computing time. This work was been supported by the SFB TR6 (DFG) within project D1. 

\bibliography{segregation}
\end{document}